\documentstyle[preprint,aps]{revtex}

\begin{document}

\title{Monte-Carlo simulations of the recombination dynamics
in porous silicon}

\author{H. Eduardo Roman}

\address{Institut f\"ur Theoretische Physik, Universit\"at Giessen,
Heinrich-Buff-Ring 16, D-35392 Giessen, Germany}

\author{and Lorenzo Pavesi\cite{add}}

\address{I.N.F.M.  and Dipartimento di Fisica, Universit\`a di Trento, via
Sommarive 14, I-38050 Povo (Trento), Italy.}

\date{received \today}

\maketitle

\begin{abstract}
A simple lattice model describing the recombination dynamics in visible
light emitting porous Silicon is presented.  In the model, each occupied
lattice site represents a Si crystal of nanometer size.  The disordered
structure of porous Silicon is modeled by modified random percolation
networks in two and three dimensions.  Both correlated (excitons) and
uncorrelated electron-hole pairs have been studied.  Radiative and
non-radiative processes as well as hopping between nearest neighbor
occupied sites are taken into account.  By means of extensive Monte-Carlo
simulations, we show that the recombination dynamics in porous Silicon is
due to a dispersive diffusion of excitons in a disordered arrangement of
interconnected Si quantum dots.  The simulated luminescence decay for the
excitons shows a stretched exponential lineshape while for uncorrelated
electron-hole pairs a power law decay is suggested.  Our results
successfully account for the recombination dynamics recently observed in
the experiments.  The present model is a prototype for a larger class of
models describing diffusion of particles in a complex disordered system.
\end{abstract}

\bigskip
Pacs: 71.55.Jv;72.20.jv;78.47.+p;61.43.Bn

\newpage
\narrowtext

\section{Introduction}

Silicon, the most studied semiconductor, is not a good light emitter,
especially in the visible range, because of its indirect band gap
transition in the infrared ($1.1$ eV).  The recently discovered optical
properties of porous forms of Silicon (p-Si) have therefore attracted
considerable interest.\cite{Canham1990,books,NATO,EMRS} In addition to the
internal nanometer sized structures which are distributed in space
according to a complex topology,\cite{Smith1992} these new materials have
similar spectral behaviors, the most prominent are a rather high
luminescence quantum efficiency and an unexpected wide range of
luminescence lifetimes (ranging from microseconds to milliseconds),
depending on temperature.

The mechanism of the luminescence emission is still unclear. Three main
models have been proposed and survived to the recent discussions:

\begin{enumerate}

\item  quantum recombination model,\cite{Canham1993,Calcott1993,Vial1992}

\item  surface state model,\cite{Koch}

\item  molecular recombination model.\cite{Prokes1992,Brandt1992,Lavine1993}
\end{enumerate}

The first two models agree on the fact that the energy spectrum of p-Si is
a result of the quantum confinement of carriers in Si nano-crystals.  Still
under debate, is the shape of the nano-crystal that has been described
either in terms of ``undulating'' quantum
wires\cite{Canham1993} or quantum dots.\cite{Vial1992} The quantum
recombination and the surface state
models differ in their predictions about the origin of the luminescence.
In the first case, the luminescence is due to confined {\it and} localized
excitons, while in the second, the luminescence is due to electron-hole
pair recombination, where the electron and the hole could be found either
in a bulk nano-crystal state ('extended' state), or being trapped into a
surface nano-crystal state ('localized' state).  The third model, the
molecular recombination model, considers that some particular molecular
species like polysilane chains (e.  g.  SiH chain\cite{Prokes1992}) or
siloxene-like rings (Si-O-H)\cite{Brandt1992} form on the surface of the Si
nano-crystals.  The luminescence is due to the carrier trapping into these
 species.

Recent experiments
\cite{Chen1992,Ookubo1992,Kondo1993,Kanemi1993,Ookubo1994,Pavesi1993,Pavesi1
995,Pavesi1995b} have provided clear evidence of anomalous
relaxation behavior of the luminescence.  The decay lineshape, for a {\it
single} observation energy, is {\it not} described by a single exponential
function.  This non-exponential behavior can be commonly described by a
stretched exponential function,\cite{Pfister1978,Klafter1986,Scher1991}
defined as
\begin{equation}
    \rm L(t) = L(0)~\exp \left[- \left(t / \tau \right) ^{\beta} \right]
    \label{decay}
\end{equation}
where L(t) is the time dependent luminescence intensity, $\tau $ is a
lifetime and $\beta \leq 1 $ a dispersion exponent.  In general, values of
$\beta <1$ correspond to the existence of a broad distribution of
lifetimes.  In some circumstances, such broad distributions may be the
result of a diffusive motion of the excited carriers.  This may be the case
in p-Si, as suggested recently in terms of trap-controlled hopping
processes.\cite{Pavesi1993} We elaborate this idea further in this work.

Motivated by these experimental results, and by the fact that little is
known about the recombination dynamics of the charge carriers in these
materials, we have initiated a detailed numerical study of the
underlying transport behavior in p-Si by means of Monte-Carlo (MC)
simulations.

In this work, we present a simple model of p-Si in which nanometer sized
crystals (nano-crystals), characterized by a distribution of radiative and
non-radiative recombination times, are assumed to be randomly placed at the
sites of percolation-like clusters defined on square and simple cubic
lattices.  Charge carriers are allowed to hop between nearest-neighbor
occupied cluster sites.  The competing effect between radiative and
non-radiative transitions in a single Si nano-crystal, as well as the
effects of geometrical constraints on hopping of carriers due to the
complex topology of percolation clusters, are discussed.

Even though our simulations are based on the quantum recombination
model,\cite{Canham1993,Calcott1993} they are consistent with other possible
recombination mechanisms because the model concentrates on the dynamics of
the recombination and not on the recombination process itself.  However, we
found that most of the data are better explained by our simulations when
the quantum recombination model is assumed.  In addition, since our model
is based on quite general assumptions about the geometrical disorder
involved, we expect that it may be applied to a larger class of physical
systems characterized by both a structural and local disorder, of which
p-Si is just one possible example.

The paper is organized as follows: in Section II the model is described,
and exact solutions obtained in the case of isolated Si nano-crystals are
discussed.  In Sections III and IV, simulation results for one-, and two-
and three-dimensional systems, respectively, are presented.  In Section V,
a discussion of the present results is presented and compared to other
results already known in the literature.  Finally, in Section VI we give
some concluding remarks.

\section{Model for a Si nanometer sized crystals}

We consider a lattice model for p-Si in which occupied and empty sites are
present.  Each occupied site of the lattice represents a Si nano-crystal.
In one dimension, all the sites are occupied by Si nano-crystals, while in
higher dimensions the occupied sites are randomly distributed in space and
interconnected to each other according to the topology of percolation-like
clusters (see Section IV).  We further assume that the physical properties,
relevant for our problem, of such Si nano-crystals are suitably described
in terms of quantum dots (QD), for which many theoretical results are
presently available.

\subsection{The size and radiative energy of a Si nano-crystal}

Each Si nano-crystal, or equivalently, each occupied site of the lattice,
is characterized by a size $d$, representing its mean diameter (typically
in the range 15 to 55 \AA \cite{sizes}).  Since the actual spatial
distribution of nano-crystal sizes is not known a priori, the sizes $d_i$
at site $i$ are chosen randomly according to a Gaussian distribution
$P(d)$.\cite{wheni} We used
$P(d)\sim(1/\sigma_d)~\exp[-(1/2\sigma_d^2)~(d-d_0)^2]$, centered at
$d_0=32.4$ \AA $\;$ and having a width $\sigma_d=0.8$ \AA.

In one dimension, this {\it local} randomness is the only source of
disorder in the model, while in higher dimensions, an additional
(geometrical) disorder, due to the complex structure of percolation-like
clusters, is also present (see Section IV).

Due to the experimental method of preparation of p-Si, however, some
correlation between the sizes $d$ may actually be present for nearby Si
nano-crystals, from which one may reasonably expect that strong local
variations in $d$ are not likely to occur.\cite{Smith1992} To take this
effect into account in a simple way, the sizes $d$ corresponding to
nearest neighbor occupied sites are not allowed to
 differ by more than a prefixed amount.  We are going
to specify this point below.

By neglecting the excitonic binding energy, the optical band-gap energy $E$
associated to a given Si nano-crystal is assumed to correspond to the
emission energy determined experimentally.  $E_i$ is related to the size
$d_i$ through the power-law relation recently suggested by
calculations performed for spherical quantum dots.\cite{Delerue} Based on
these calculations, we assume
$E_i= E_0 (d_0/d_i)^n$, with $n\cong 1.39$, where
$E_0=1.86$ eV is the transition energy for a Si nano-crystal of size $d_0$.
The corresponding distribution for $E$ is then of the form,
\begin{equation}
\label{Gauss}
P(E)\sim {1\over\sigma_d}~{1\over E^{1+1/n}}~
\exp[-(d_0^2/2\sigma_d^2)~( (E_0/E)^{1/n}-1)^2]
\end{equation}
centered around $E_0$ and having a width at half-maximum of about 0.16 eV
(see Fig.  \ref{f-Gaus}).

It is instructive to obtain estimates of the total density of states
$\rho(\epsilon)$ resulting from these distributions.  For simplicity, an
ensemble of quantum dots of cubic shape and side $L_x=L_y=L_z=d$ is
considered.\cite{Yoffe1993}  The (confined) electronic states have energies
$\epsilon(n_{x,y,z})=\epsilon_0 (n_x+n_y+n_z)$, where
$\epsilon_0=\hbar^2\pi/(2\mu d^2)$, $n_x$, $n_y$ and $n_z$ are the
principal quantum numbers for motion in the $x, y$ and $z$ directions,
$n_{x,y,z}\ge 1$ and $\mu$ is the effective mass.  The transition energy
$E$ can be related to the ground state energy of the QD by
$E=3\epsilon_0+E_{\rm Bulk}$, where $E_{\rm Bulk}=1.15$ eV is the bulk-Si
indirect gap.  The density of states for a single quantum dot of size $d$
with an energy barrier of height V(QD) is then\cite{Yoffe1993}
\begin{equation}
\label{Densd}
\rho_d(\epsilon)\cong {2 V({\rm QD})\over d^3}\sum_{n_{x,y,z}}~
\delta(\epsilon-\epsilon(n_{x,y,z}) - E_{\rm Bulk}),
\end{equation}
where $\epsilon$ is an energy and the total density of states is
\begin{equation}
\label{Denst}
\rho(\epsilon)=\sum_d ~P(d)~\rho_d(\epsilon).
\end{equation}
This quantity is plotted in Fig. \ref{f-Gaus} assuming a Gaussian
broadening of the single QD state of 0.01 eV.
The density-of-states of an ordered ensemble of QDs should reveal a series
of distinct peaks corresponding to the different confined states
$\epsilon(n_{x,y,z})$ in the QD.  The random Gaussian distribution of
sizes, i.  e.  E, smears out the discrete peaks and yields an approximately
exponential behavior of the form $\rho(\epsilon)\sim
\exp(\epsilon/\epsilon_a)$, with $\epsilon_a\cong 0.35$ eV, at large
$\epsilon$.

To deal with the above mentioned spatial correlations in $d$, the
difference in energy between nearest-neighbor sites $i$ and $j$, $\Delta
E_{ij}=E_j-E_i$, is allowed to take absolute values smaller than a given
cutoff $E_{\rm cut}$.
This is the criterion employed to accept a given value of $d_j$, with
respect to the previous value at site $i$, from the Gaussian distribution.
In dimensions higher than one, this criterion is applied simultaneously
with the growing process (see Section IV).  For convenience, we consider
variations in energy, rather than variations in size to select the values
of the local quantities $E$ and $d$.  The results are not sensitive to this
choice. Typical calculations have been performed for
values in the range $10^{-4}<E_{\rm cut}<0.4$ eV (see Section III).
 $E_{\rm cut}$ is the first parameter in the model.

\subsection{The radiative and non-radiative recombination times}

Within each Si nano-crystal, both radiative and non-radiative recombination
processes occur when excited electron-hole pairs are present.  Both
processes depend on the size $d$ of the nano-crystal.

The emission process is assumed to be due to excitonic recombinations in a
QD.  In this case, it was shown that the radiative recombination time,
$\tau_{\rm rad}$, results from the thermal balance between the occupation
of the exchange splitted singlet and triplet excitonic
states,\cite{Calcott1993}
\begin{equation}
\label{taurad}
\tau_{\rm rad}(E,T)=\tau_{\rm tripl}\left[
{1+(1/3)\exp(-\Delta E_x/k_B T)\over
1+(1/3)(\tau_{\rm tripl}/\tau_{\rm sing})~
\exp(-\Delta E_x/k_B T)}\right].
\end{equation}
$\tau_{\rm sing}$ (the radiative lifetime for the singlet excitonic state)
is obtained from the values calculated in Ref.  \onlinecite{Hybertsen},
\begin{equation}
\label{tausin}
\tau_{\rm sing}(E)=\tau_{\rm sing}^{(0)}
\left( {E_{\rm Bulk}\over E-E_{\rm Bulk}} \right)^{{\textstyle n_s}}
\end{equation}
where
$n_s=1.5$, and $\tau_{\rm sing}^{(0)}$ is the second parameter in the
model. At the present, no theoretical calculations are available for the
radiative lifetime of the triplet excitonic state,
$\tau_{\rm tripl}$. Thus,
$\tau_{\rm tripl}$ is estimated from
the experimental data reported in Ref. \onlinecite{Calcott1993},
\begin{equation}
\label{tautri}
\tau_{\rm tripl}(E)=\tau_{\rm tripl}^{(0)}~ \left[ \tau_{\rm
sing}(E)\right]^{{\textstyle
\alpha_s}}
\end{equation}
where $\tau_{\rm tripl}^{(0)}=2368~\mu{\rm s}$, and $\alpha_s=0.307$.

Finally, the quantum confinement enhanced exchange energy $\Delta
E_x=E_x^{Si}~(d_x/d)^3$, where $E_x^{Si}=3.165\times 10^{-4}$ eV is the
exchange energy in crystalline Si and
$d_x=43$ \AA, is obtained from the calculations in Ref.
\onlinecite{Fishman}.

Non-radiative recombinations have been considered as multiphonon
transition processes for trapping on a deep center,\cite{Ridley}
for which the non-radiative recombination time, $\tau_{\rm nr}$,
is given by
\begin{equation}
\label{taunr}
\tau_{\rm nr}(T)=\tau_{\rm nr}^{(0)} { e^{2 n_T S} \over (n_T+1)^p}
\end{equation}
where $n_T=[\exp(\hbar\omega_{ph}/k_B T)-1]^{-1}$ is the density of phonons
with energy $\hbar\omega_{ph}$, and $S$ and $p$ are related to the details
of the capturing center.  Here we have used nano-crystal size independent
quantities, with $\hbar\omega_{ph}/k_B =800$ K, $S=1$, and $p=25$.  If
bulk-Si optical phonons are involved, $\hbar\omega_{ph}/k_B \simeq 1100$ K;
lower values can be justified by the typical softening of the phonon modes
in a confined system on a localized center.  $\tau_{\rm nr}^{(0)}$ is the
third parameter in the model.

It should be emphasized that the actual nature and origin of non-radiative
phonon transition processes in p-Si nano-crystals remain to be understood.
Here, such processes are included because they certainly play a very
important role in the recombination dynamics of
carriers.\cite{Vial1992,Koch} The form for $\tau_{\rm nr}$ in Eq.
\ref{taunr} can be only considered as a first approximation to the actual
form of the non-radiative recombination time, yet it is expected to
describe the physics involved correctly, at least in a qualitative fashion.
In this work, the same $\tau_{\rm nr}$ has been employed for all the Si
nano-crystals.

\subsection{The hopping rates between Si nano-crystals}

In addition to the internal structure of Si nano-crystals, hopping of
electrons, holes and excitons, may actually occur between nearby
nano-crystals.  To describe the diffusion of carriers in p-Si, we consider
hopping processes only between nearest neighbor (n.n.) occupied sites of
the lattice for two different models.  In the first model, only one type of
carrier hops, representing the exciton, while in the second, two different
types of carriers can hop, representing the electron and hole.  If the
electron and hole (e-h) motions are strongly correlated, e.g.  if the hole
closely follows the electron as it moves in the system, one has essentially
the first model.  As we will see in Sections III and IV, a qualitatively
different behavior is obtained when e-h hops are completely uncorrelated.
In both models, correlated and uncorrelated e-h pairs, the on site radiative
recombination is assumed to be excitonic.

The transition rates for a carrier of type $x$ ($x\equiv ex$, representing
the exciton, $x\equiv e$, the electron, and $x\equiv h$, the hole)
from the site $i$ to its nearest-neighbor
occupied site $j$ is defined as
\begin{equation}
\label{hopm}
P_{ij}^x= {1\over \tau_{\rm hop}^x}  \qquad  {\rm when} \qquad
                                         E_j-E_i\le 0
\end{equation}
and
\begin{equation}
\label{hopp}
P_{ij}^x= {1\over \tau_{\rm hop}^x}~\exp(-f_x \Delta E_{ij}/k_B T)
                              \qquad  {\rm when} \qquad   E_j-E_i\ge 0
\end{equation}
where $\Delta E_{ij}=E_j-E_i$.  When the site $j$ is not occupied, as it
may occur in two- and three-dimensional systems, then $P_{ij}^x=0$.  Here,
$0<f_x\le 1$ is an additional parameter related to the band structure of
nano-crystals, and $\tau_{\rm hop}^x$ is a characteristic time for
tunneling of $x-$carriers between nearby (n.n.) Si nano-crystals, which is
temperature independent.  In Eqs.  \ref{hopm} and \ref{hopp}, no dependence
on the intersite distance, $r_{ij}$, between n.n.  Si nano-crystals is
explicitly shown.  In a more refined version of the present model, one
could consider in Eqs.  \ref{hopm} and \ref{hopp} a dependence on $r_{ij}$
of the form
\begin{equation}
        {1\over \tau_{\rm hop}^x}= \nu_x \exp(-\gamma_{ij}r_{ij})
\end{equation}
where $\nu_x$ is an attempting frequency and $\gamma_{ij}$ is related to
the energy barrier which separates the QDs at site $i$ and $j$.  The
quantities $\gamma_{ij}$ and $r_{ij}$ may then be considered as random
variables.  For simplicity, we use $\tau_{\rm hop}^x=\tau_{\rm hop}$ here,
independently of the type of carrier.  Thus, the dependence of hopping
rates on $x$ is only contained in the factor $f_x$.

In our calculations, we have arbitrarily chosen $f_e=1/3$, $f_h=2/3$ and
$f_{ex}=0.4$ in order to take into account the distribution of the
energy-gap discontinuity, ${\rm E-E_{Bulk}}$, among the valence and
conduction bands.  Extensive Monte-Carlo simulations have been done by
using the same set of parameters but for different $f_{ex}$.  The effect of
variing $f_{ex}$ is to change the barrier for hopping, $f_{ex} \Delta
E_{ij}$, which is mostly relevant at low temperatures.  By increasing
$f_{ex}$, a decrease in the luminescence decay rate occurs, but essentially
no changes in the luminescence decay lineshapes are observed, i.  e.  in
Eq.  \ref{decay} $\tau$ increases and $\beta$ remains constant.  The mean
distances over which the excitons move before recombination are reduced as
the effective barrier, $f_{ex} \Delta E_{ij}$, is increased.  However, for
$f_{ex}$ in the interval $\left[ 0.2-0.6 \right]$ the numerical results are
qualitatively not affected and quantitatively they change by less than 10
$\%$.

Before going on with the discussion of the results, we briefly
summarize the free parameters entering the model:

\noindent
(1) $E_{\rm cut}$, absolute energy difference cut-off of nearby occupied
sites in the lattice.

\noindent
(2) $\tau_{\rm sing}^{(0)}$, prefactor for the singlet radiative
recombination time (Eq. \ref{tausin}).

\noindent
(3) $\tau_{\rm nr}^{(0)}$, prefactor for the non-radiative
recombination time (Eq. \ref{taunr}).

\noindent
(4) $\tau_{\rm hop}$, prefactor for the tunneling of carriers
between nearest-neighbor occupied sites (Eq. \ref{hopm} and \ref{hopp}).

Table I reports the values of all the parameters used in our
simulations for 1D, 2D and 3D systems whose results are presented in the
following.

\subsection{The case of isolated Si nano-crystals}

Let us consider the case in which no hopping between n.n.  Si nano-crystals
occurs in the system.  This limiting situation may well describe p-Si
samples in which Si nano-crystals are efficiently passivated and surrounded
by thick surface oxide layers.  This case represents an ensemble of
isolated QDs, each characterized by its own size $d$ and transition energy
$E$.  To obtain the luminescence spectrum of the whole ensemble as a
function of time, for different observation energies and temperatures, the
rate equation describing the time evolution of the density $N(t)$ for a
single carrier at a given Si nano-crystal (site) is solved.  Such a rate
equation reads,
\begin{equation}
\label{rate}
{d N(t)\over dt} = - {N\over \tau_{\rm rad}} -
                     {N\over \tau_{\rm nr} }
\end{equation}
and the exact solution is
\begin{equation}
\label{taux}
N(t) =  N(0)~\exp(-t/\tau)   \qquad   {\rm with}   \qquad
          {1\over \tau}={1\over \tau_{\rm rad}} +
                        {1\over \tau_{\rm nr}}.
\end{equation}
Typical behaviors of the different times entering in Eq. \ref{taux} as a
function of temperature, for fixed energy $E$, are shown in Fig.
\ref{f-taus}.

The luminescence $L(t)$ from a single Si nano-crystal is just
$L(t)=N(t)/\tau_{\rm rad}$, i.e.
\begin{equation}
\label{Lumis}
L(t) =  L(0)~\exp(-t/\tau)
\end{equation}
yielding simple exponential decay, i.e.  $\beta=1$ (cfr.  Eq.
\ref{decay}).  As pointed out in Ref.  \onlinecite{Vial1992}, the
luminescence decay time is essentially determined by non-radiative
recombinations at room temperature.  According to Eq.  \ref{taux} and Fig.
\ref{f-taus}, non-radiative recombinations play a main role also at low
temperatures when the radiative lifetime, dominated by the triplet
recombination lifetime, becomes very large.  In the intermediate
temperature range, when the quantum efficiency also increases, the singlet
lifetime may dominate the recombination lifetime, if it becomes
sufficiently small and thermal quenching of the phonon population takes
place, decreasing the capture cross-section to the non-radiative
recombination centers.  This is case 3 in Fig.  \ref{f-taus}.  At
sufficiently high temperatures, non-radiative recombinations dominate
again.

The total radiated intensity
$I(E,T)=\int_0^\infty dt~L(t)=N(0)~\tau/\tau_{\rm rad}$,
is related to the corresponding quantum efficiency, $\epsilon(E,T)$
defined as
\begin{equation}
\label{QE}
\epsilon(E,T) \equiv {I(E,T)\over N(0)} =
{\tau_{\rm nr} \over \tau_{\rm nr}+\tau_{\rm rad}}.
\end{equation}
For illustration, the quantity $\epsilon(E,T) P(E)$, representing the
quantum efficiency of the ensemble of QDs described by Eq.  \ref{Gauss} for
an energy $E$, is plotted for several temperatures versus energy in Fig.
\ref{f-qee}.  The total quantum efficiency of the ensemble, $\epsilon(T)$,
is just the sum of Eq.  \ref{QE} over the total number of Si nano-crystals
in the system, i.e.  $\epsilon(T)=\sum_E~P(E)~\epsilon(E,T)$.  This
quantity is shown in Fig.  \ref{f-qex}.  At high temperatures, we find an
exponential behavior $\epsilon(T)\sim \exp(-T/T_0)$, where $T_0\cong 59$ K
(see cases (1) and (2) in Fig.  \ref{f-qex}), in good agreement with recent
experimental results (see e.g.  Ref.  \onlinecite{Rosenbauer}).  The value
$\hbar\omega_{ph}/k_B =800$ K and $p=25$ (see Section II.B), were chosen
such that $\epsilon(T)$ displays a maximum in the range $100$ K$<T<200$ K
and a value of about 10 \% at room temperature, as is experimentally
observed.\cite{Vial1992}

Finally, the absorption coefficient $\alpha(E)$ of the ensemble can be
obtained from the relation
\begin{equation}\label{alp}
\alpha(E,T) \simeq{{\rm const}\over E}~\vert M\vert^2~\rho(E),
\end{equation}
where $\vert M\vert$ is the dipole matrix element at the band-edge
describing radiative transitions in the QD and $\rho(E)$ is the total
density of states.  $\vert M\vert$ is related to the oscillator strength of
the transition and hence to the radiative lifetime, i.e.  $\vert
M\vert^2\sim \tau^{-1}_{\rm rad}(E,T)$.\cite{Hybertsen} Using the result
obtained in Section II.A for $\rho(E)$, Eq.  \ref{Denst}, and the
expression Eq.  \ref{taurad} for $\tau_{\rm rad}$, we obtain the absorption
coefficient of the ensemble as displayed, for various temperatures, in Fig.
\ref{f-abs}.  Remarkably, this very simple model is able to reproduce the
main features of the measured absorption coefficient,\cite{Koch} i.e.  a
quite sharp edge at about 1.6 eV and an increasing exponential shape at
high energies with a slope which depends on the temperature.  An
exponential fit at high energies, $\alpha(E)\sim \exp(E/E_0)$, of the
curves shown in the figure yields for the energy slopes: $E_0=0.55$ eV for
$T=30$ K, 0.34 eV for $T=100$ K, 0.29 eV for $T=200$ K and 0.28 eV for
$T=300$ K.  The temperature dependence of $\alpha (E)$ is due to the
$\tau_{\rm  rad}$ term in Eq.  \ref{alp}.  The exponential dependence of the
absorption coefficient has been repetitively observed in
p-Si\cite{Rosenbauer,Bustarret} and attributed to an Urbach-like edge as in
amorphous Silicon.  The origin of the Urbach-tail is not precisely known so
far but it is believed to be due to disorder.  In the present model, it
results from the overlap of the 0-dimensional densities of states
associated to the QDs randomly arranged in space.

\subsection{Rules for the Monte-Carlo simulations}

In the case in which hopping of carriers between nearby Si nano-crystals
takes place, the rate equation for $N(t)$, Eq. \ref{rate}, is modified to
\begin{equation}
\label{rateh}
{d N_i(t)\over dt} = - {N_i\over \tau_{\rm rad}} -
                       {N_i\over \tau_{\rm nr}} -
                       \sum_j~P_{ij}~N_i(t) + \sum_j~P_{ji}~N_j(t)
\end{equation}
yielding a system of coupled differential equations for the different
occupied sites $i$.  For simplicity, we have omitted the index $x$,
denoting the type of carrier, from Eq.  \ref{rateh}.  The third term in Eq.
\ref{rateh} represents the outgoing particles from site $i$ (loss term),
while the last term the incoming particles to site $i$ from the neighbor
sites $j$ (feeding term).  In one dimension, the number of nearest neighbor
sites $n=2$, and because of structural disorder, in two dimensions $1\le
n\le 4$, and in three dimensions $1\le n\le 6$.  The system in Eq.
\ref{rateh} is conveniently solved by Monte-Carlo (MC) simulations when the
total number of effectively coupled occupied sites $i$ becomes large.  In
particular, certain constraints for the occurrence of recombination
processes, such as the simultaneous presence of an electron and a hole at a
given site, can be easily implemented with the MC method.

Before discussing the MC-rules, we need to determine the unit of time,
denoted as $\tau_0$, which should be sufficiently small such that
faster transition events are well described. Once the
transition times $\tau$ have all been determined in the system, one
can take $\tau_0=\tau_{\rm fast}/n_t$, where $\tau_{\rm fast}$
is the smallest transition time in the ensemble and $n_t>1$.
We have used $n_t=10$ in our simulations. Notice that the time $t$
becomes now a discrete variable, i.e.
$t=n \tau_0$, where $n\ge1$ denotes the $n$th MC-step.
The total elapsed time for each MC-step is just $\tau_0$.

We can now discuss the MC-rules for the present model.  Since we are
interested only in the case of very low carrier density, i.e.  different
electron-hole pairs in the system do not see each other (no interaction
effects or Auger recombinations), we study the time evolution (trajectory)
of a $single$ electron-hole pair (either as an exciton or as two
independent particles), which is initially located ($t=0$) at the center of
the lattice.  Averages are then performed over many trajectories and
different realizations of disorder.

During the $n$th MC-step, a particle at site $i$ may undergo one
of the following four different processes:

\noindent
(1) decaying radiatively (annihilate) with a probability
$p_{\rm rad}=\tau_0/\tau_{\rm rad}$,

\noindent
(2) decaying non-radiatively (annihilate) with a probability
$p_{\rm nr}=\tau_0/\tau_{\rm nr}$,

\noindent
(3) hopping to a n.n. site $j$ with a probability
$p_{ij}=\tau_0~P_{ij}$, or

\noindent
(4) remaining at site $i$ with a probability
$p_{\rm sit}=1-p_{\rm rad}-p_{\rm nr}-\sum_j~p_{ij}$.

In the case in which an electron and a hole are considered,
a decay can only take place when both carriers are located
at the same site.

According to our definition of $\tau_0$, all the above probabilities are
smaller than one as required.  To decide which event will take place for a
given particle at site $i$, we generate a random number $r$, uniformly
distributed in the range $0<r<1$.  Then, we evaluate the partial sums over
the probabilities, i.e.  $\sum_{k=1}^m~p_k=S_m$, where the index $k=1$
represents a radiative recombination event, $k=2$ a non-radiative one,
etc., and $m\ge1$.  The successful event is the one for which $S_m>r$ for
the first time.  For example, consider a 1D system and let $p_{\rm
rad}=p_{\rm nr}=0.2$ (equally likely radiative and non-radiative
recombinations) and $p_{i(i-1)}=p_{i(i+1)}=0.1$ (i.e.  equal hopping rates
to two n.n.  sites).  Then, if $r=0.55$, the particle will undergo the
fourth possibility, i.e.  it will hop to the second n.n.  site.  If,
however, $r>0.6$, the particle will remain sit at its present site.

A single MC-step is completed when the above discussed procedure has been
applied to each particle in the system (either the exciton, or the electron
and hole), and the time $t$ is increased by $\tau_0$, $t=t+\tau_0$.  At the
$n$th MC-step, corresponding to time $t=n \tau_0$, the luminescence of the
system is simply obtained by counting the total number of radiative events
during the $n$th MC-step.  Since we are also interested in the energy
dependence of the luminescence, those radiative events originated at sites
characterized by transition energies $E'$ close to the observation energy
$E$ (i.e.  $\vert E'-E\vert>\delta E$) are recorded separately.  In our
simulations we used $\delta E=0.01$ eV.

We proceed with the numerical results obtained for one-dimensional
lattices.

\section{Results for one-dimensional lattices}

At very high porosities, p-Si is characterized by a filamentary structure
having a quasi-one dimensional character on small length
scales.\cite{Teschke1993} Thus, we start our discussion about transport
(diffusion) of carriers in p-Si on a simplified one-dimensional
model. Preliminary results with the set of parameters named case (3) in
Figs. \ref{f-taus} and \ref{f-qex} are reported in Ref.
\onlinecite{Pavesi1995}.

When the carriers are allowed to hop between n.n.  sites of the linear
lattice, the diffusive motion takes place in a highly irregular energy
landscape.  This feature of the model is illustrated in Fig.
\ref{f-1dprofil}, where the local transition energies $E$ are plotted as a
function of position on the one-dimensional lattice.  In the figure, three
cases are considered for selected values of the parameter $E_{\rm cut}$
(see Section II.A).  Although local variations in energy can be smoothed
out when $E_{\rm cut}$ becomes sufficiently small, on large length scales
appreciable variations of $E$ will always occur (see Fig.
\ref{f-1dprofil}).  This has important consequences when considering
hopping processes, because carriers must eventually overcome an effective
large barrier to diffuse out of a region of local minimum energy.
According to our definition of the transition rates in Eq.  \ref{hopp},
thermally activated hopping against such large scales effective barriers
will be strongly hindered at sufficiently low temperatures, even in the
case of low $E_{\rm cut}$ values.  The low energy sites with low $p_{\rm
rad}$ and high energy barriers $\Delta E_{ij}$ act as {\it temporary
traps}.  Excitons which relax to these sites are temporary trapped into
them and have to wait long times before being released.

To quantify the effect of disorder on diffusion we calculate the mean
square displacement of the carrier as a function of time, $R^2\equiv
\langle r^2(t)\rangle-\langle {\vec r}(t)\rangle^2$, and investigate its
temperature dependence.  ${\vec r}(t)$ is the displacement with respect to
the initial position, which is assumed to be the origin.  To study the
effect of the dynamics of carriers on the time dependence of $L(t)$, we
have solved Eq.  \ref{rateh} for the values given in the Table I.
Since in experiments all sites of the system are homogeneously excited at
$t=0$, including those with energy $E$ close to the observation energy
$E_{\rm obs}$, initially, the electron-hole pair is placed at a site (the
center of the lattice) characterized by an energy $E\cong E_{\rm obs}$.

In the case of excitons, the behavior of the luminescence is shown in Fig.
\ref{f-1dexc} for four different temperatures and observation energy
$E_{\rm obs}=1.86$ eV.  Two remarks are evident: i) the decay departs from
the simple exponential decay obtained for isolated QDs (corresponding to
$\tau_{\rm hop}=\infty$), and ii) the time scale of the decay is strongly
temperature dependent.  The last item is due to the thermal population of
the singlet state (fast radiative recombination time) and to the onset of
efficient non-radiative decay, especially at the highest temperature.  The
lineshape analysis shown in Fig.  \ref{f-1dfit} evidences that a stretched
exponential function is required to fit the decay.  The single exponential
or the double exponential functions are able to fit the decay only on a
limited time range.

The stretched exponential decay reflects the fact
that a time delay between recombination events at a given QD may occur when
the exciton can diffuse out of the QD.  Such time delay can persist on
large time scales since the exciton may either diffuse far away from the
site at which the radiative event is expected, or may become temporarily
{\it trapped} at a nearby site with a lower transition energy.  When the
observed radiative events are collected from sites with similar transition
energies $E$, the random local environment around each of these sites leads
to a distribution of recombination times, which are in general different
than the single value $\tau_{\rm rad}(E,T)$ for that energy $E$.  This {\it
dispersion} of recombination times causes the stretched exponential
behavior observed in the model.

In the case of uncorrelated electron and hole motion, a quite different
behavior is observed (see Fig.  \ref{f-1deh}).  At high temperatures,
and intermediate times, an approximate power-law decay of the form
\begin{equation}
\label{Lpower}
L(t)\sim t^{-\alpha}
\end{equation}
takes place. The range of times
over which the power-law decay seems to occur grows as the temperature is
raised. Here, the uncorrelated motion of the electron and hole leads
to a dramatic slowing down of the luminescence decay, since the two
carriers are no longer constrained to be all the time simultaneously
at the same site. Similar lineshapes have been observed for amorphous
hydrogenated Silicon.\cite{Noolandi1980} In this system, the luminescence
is explained as due to geminate recombination between e-h pair localized
in band tail states.\cite{Street1974}

We have performed extensive MC simulations for different temperatures.  In
the case of excitons, the decay lineshapes are fitted by a stretched
exponential function for all the temperatures considered.  A statistical
weigth of the simulation points has been used to obtain the best fit
function.  The results of the fits are summarized in the upper panel of
Fig.  \ref{f-taubet}.  Typical errors for $\tau$ of 10 $\%$ and $\beta$ of
0.02 are expected, due to statistical fluctuations and goodness of the
fits.  The lifetime $\tau$ decreases significantly as the temperature is
raised, while the dispersion exponent $\beta$ increases and tends to unity
at high temperatures.

We have also calculated the quantum efficiency $\epsilon(E,T)$ as a
function of $E$ for a fixed temperature, displaying a maximum around
$E=1.86$ eV and found that hopping of carriers does not affect sensitively
$\epsilon(E,T)$.  This is clear because $\epsilon(E,T)$ is essentially
determined by the on-site recombination dynamics.  Independently of the
dispersive motion, all the carriers recombine radiatively or non-radiatively
after a sufficiently long time.  The diffusion only influences the
recombination dynamics by introducing long lived recombinations.

In Fig.  \ref{f-1dmsd}, we show the values of $R^2(t)$ for different
temperatures.  At intermediate and high temperatures, an anomalous behavior
of the form $R^2(t)\sim t^{2/d_w}$ with $d_w\ge 2$, takes place.  Such
anomalies in $R^2(t)$ are typical of diffusion on fractals, where the
presence of dangling ends and loops on all length scales in the fractal
slows down the diffusion process on all time scales.  On fractals, however,
the diffusion exponent $d_w$ is temperature independent,\cite{Bunde} while
here $d_w$ turns out to depend strongly on temperature (see Fig.
\ref{f-dw}).  The anomalies in $R^2$ are due to the random distribution of
hopping rates, which is known to yield values of $d_w>2$ and dependent on
temperature (see e.g.  Ref.  \onlinecite{Silver1982}).  At low
temperatures, however, the time dependence of $R^2(t)$ departs from a power
law.

At sufficiently low temperatures, indeed, the present model is expected to
display transport behavior similar to diffusion in the presence of random
fields (Sinai model), see Refs.  \onlinecite{Sinai} and
\onlinecite{Roman1}.  To show this, we have drawn in Fig.  \ref{f-sinarf} a
small section of a typical energy landscape versus position.  The arrows
represent the local {\it fields} felt by the carrier which is proportional
to the energy difference $\Delta E_{ij}$ between the two n.n.  sites $i$
and $j$.  Due to the assumed random distribution of nano-crystal sizes in
the system, the local fields are randomly oriented, similarly as in the
Sinai model.  Diffusion of carriers in the system is biased by these random
fields and becomes ultra-anomalously slow.  Of course, in the present model
the energy barriers (and also the potential valleys) can not grow
indefinitely as in the Sinai case.  However, when the temperature is
sufficiently low, diffusion is strongly hindered and the carriers can only
explore small length scales since the effect of the random fields becomes
dominant.  In these circumstances, a logarithmic time dependence is
expected
\begin{equation}
\label{logt}
R^2(t) \sim (\log t)^4
\end{equation}
at long times.  Results obtained for the case of excitons are plotted in
Fig.  \ref{f-1dlogt}.  These suggest that for large times and low
temperatures, our model displays logarithmic time dependencies of the form
predicted by Eq.  \ref{logt}.  To our knowledge, this is the first time
that such logarithmic time dependencies of $R^2(t)$ are predicted for a
model aimed to describe a disordered system such as porous Silicon.
Actually, the value of the exponent describing the logarithmic time
dependence turns out to depend slightly on temperature, and the value of
four is observed only for $T\cong 30$ K.  At lower $T$, $R$ tends to
extrapolate to a constant value because non-radiative annihilation events
become dominant, and the carriers can only explore a finite spatial extent
asymptotically.  A similar trend is also observed for electrons and holes
in the case of uncorrelated electron-hole pairs.

\section{Two- and three-dimensional models and results}

The actual geometrical structure of porous Silicon is not known accurately.
For many purposes, however, one may hope to capture the essential features
of such complex structures by modeling them with simple, yet non trivial,
percolation-like clusters.\cite{Bunde} The clusters are
generated by using a modified version of the well-known growth algorithm
employed for percolation clusters (see e.g. Ref. \onlinecite{Bunde}), which is
adapted here to our present purposes.

For simplicity we consider a square lattice in two dimensions, and a simple
cubic one in three dimensions.  The linear size of the lattice is denoted
by $\ell$.  The growing process starts at the seed, which is located at the
center of the otherwise empty lattice.  The growth proceeds according to
the following rules.  A nearest neighbor site $j$ of the seed can be
occupied with probability $p_n$, with the index $n$ indicating the number
of occupied sites nearest to the growing site $j$.  Initially, $n=1$.  If
the site $j$ is not occupied, i.e.  if it does not become part of the
cluster, it is blocked and cannot be occupied later.  The process continues
from the last occupied sites and now values $n>1$ may occur.

For standard percolation clusters, one takes $p_n=p$ independently of $n$.
When $p<p_{\rm crit}$, $p_{\rm crit}\cong 0.593$ (two dimensions), and
$p_{\rm crit}\cong 0.312$ (three dimensions), only finite clusters can
grow, while infinite clusters develop when $p>p_{\rm crit}$.  When
$p=p_{\rm crit}$, large percolation clusters may grow which are fractal
(with fractal dimension $d_f\cong1.896$ in two dimensions, and
$d_f\cong2.5$ in three dimensions) on large length scales.  Finite clusters
and the infinite cluster above $p_{\rm crit}$ are also fractal (with the
same fractal dimension $d_f$) for length scales smaller than the
correlation length.  The latter diverges at $p=p_{\rm crit}$.

In our model, we take $p_1>p_2$ and $p_n=0$ when $n>2$.  In the following,
we consider the cases, $p_1=0.65$ and $p_2=0.15$ in two dimensions, and
$p_1=0.45$ and $p_2=0.17$ in three dimensions.

We notice that when $p_1\to1$, and $p_2\to0$, the clusters tend to grow
linearly, while for relatively large $p_2$, compact clusters can be grown.
By varying both $p_1$ and $p_2$, a variety of structures can be obtained
which are suitable for our present purposes.  A measure of the {\it
porosity} of the cluster is simply given by the ratio between empty sites
and occupied sites.  Typical examples are shown in Fig.  \ref{f-poros3} for
three dimensional clusters, and in Ref.  \onlinecite{Pavesi1995b} for
two-dimensional clusters.  The clusters are uniform on large length scales,
but still display spatial fluctuations on small length scales as the
infinite cluster above $p_{\rm crit}$.\cite{Pavesi1995b} Thus, the
intrinsic fractal character of percolation is common to the clusters shown
in Fig.  \ref{f-poros3} too.  It is possible to construct highly
constrained and filamentary clusters (like that in Fig.  \ref{f-poros3}(a))
or more compact clusters (like that in Fig.  \ref{f-poros3}(b)).

We have studied the dynamical behavior of carriers on these clusters,
following the same MC-rules described above.  The question now is how the
extra degrees of freedom in space, resulting from the two- or
three-dimensional topology of the clusters, modify the time decay of
$L(t)$.  Preliminary results for two-dimensional systems have been reported
in Ref.  \onlinecite{Ceschini1994}.  Results for three-dimensional systems
are reported in Fig.  \ref{f-lumi3} for the luminescence
decay.\cite{Pavesi1995b} For excitons and uncorrelated electron-hole pairs,
the lineshape is a stretched exponential and a power law, respectively.
Hence the decay lineshape is not modified by the increased dimensionality.
Results of stretched exponential fits to the luminescence decay in two or
three dimensions, for the exciton model, are reported in Fig.
\ref{f-taubet} as a function of temperature.

The three-dimensional topology has important consequences on the decay of
the luminescence.  The $\tau $ and $\beta$ values are lower, indicating
that the role of hopping processes in three dimensions is more important
than in lower dimensions.  While the one-dimensional model (and also the
two-dimensional one) predicts values of $\beta$ close to unity already at
room temperature, in three dimensions, in contrast, the theoretical values
are consistent with the experimental results for $\beta$, which typically
saturate at values $\beta\cong 0.7-0.8$.\cite{Pavesi1993,Pavesi1995} Thus,
values of $\beta<1$ at high temperatures, and intermediate porosities, can
be explained by our model as a result of the interplay between a complex
conducting matrix (representing the topology of p-Si), and an additional
local disorder due to the distribution of nano-crystal sizes in the system.

A comparison of the temperature dependence of $\tau$ and $\beta$ for two
different sets of parameters (but for the same growth parameters $p_1$,
$p_2$ mentioned above) is reported in Fig.  \ref{f-compare3D}.  The effect
of variing $\tau_{\rm nr}$ is evident in $\beta$: the larger the $\tau_{\rm
nr}$ the lower is $\beta$.  The role of temporary traps, which tend to
reduce $\beta$ and are effective at low temperatures, is enhanced when
$\tau_{\rm nr}$ is increased due to a relative increase of the role of
exciton diffusion ($\sim \tau_{\rm nr}/\tau_{\rm hop}$).

The dynamics of carriers is influenced by the parameters $p_1$ and p$_2$
used to construct the clusters.  The more the cluster is filamentary (see
e.g.  case (a) in Fig.  \ref{f-poros3}), the more the excitons are
constrained to move along one direction, and the larger are $\tau$ and
$\beta$.  Then, the dynamical behavior of carriers should have essentially
a one-dimensional character, at least for sufficiently low temperatures
where transport along the structure is strongly hindered.  Thus, we expect
to observe, in this regime, the same logarithmic time dependence of
$R^2(t)$ as in the one-dimensional model, Eq.  \ref{logt}, at short times.

For arbitrary porosities and not too low temperatures, we find anomalous
behavior of $R^2(t)\sim t^{2/d_w}$, at intermediate times (Fig.
\ref{f-3dlogt}).  Here, $d_w>2$ and increases as the temperature decreases,
more than in the one-dimensional case (Fig.  \ref{f-dw}).  Eventually, a
logarithmic time dependence of $R^2(t)$ is observed for sufficiently low
$T$.  In three dimensions, we have found clear evidence for the behavior
\begin{equation}
\label{logt2}
R^2\sim (\log t)^2
\end{equation}
at long times (Fig.  \ref{f-logt2}), in contrast to the fourth-power result
valid for the Sinai model.  This new behavior represented by Eq.
\ref{logt2} has been predicted for the case of biased diffusion on random
fractals in two dimensions.\cite{Roman1} As we can see, a similar behavior
can be expected also in three dimensions.  This feature remains to be
studied theoretically.

\section{Discussion}

According to the present model, the photoluminescence decay, L(t), is
determined, in the case of excitons, by the recombination and the escape
rates from the isolated QD.  In the case of uncorrelated electron-hole
pairs, L(t) is determined by the diffusion of the electron and the hole,
because both need to be on the same QD to recombine radiatively.

In the case of excitons, the disordered environments of the QDs, from which the
luminescence is observed, cause a distribution of waiting times for hopping
and/or release times for activated emission from the temporary traps,
and the luminescence displays a stretched exponential decay.
In the case of uncorrelated electron-hole pairs,
the main role is played by the encounter probability of the electron
and the hole to be on the same site.  This is determined by the diffusion of
the
slower moving particle, i.e. the hole,
and the time dependence of the recombination
rate becomes a power-law,\cite{Scher1991}
which is then reflected in the luminescence decay.

For all spatial dimensions $D$, the increase in $\beta$ is almost linear
for 10 $<$T$<$60 K, has a plateau for 60 $<$T $\leq$ 200 K and then
increases rapidly to 1 at higher temperatures.  This behavior corresponds
to the following three typical regimes: a first one in which the temporary
traps play a role (low temperatures), a second one in which hopping
dominates and diffusion is restricted only by the p-Si network geometry
(intermediate temperatures), and a third one, in which the rapid on-site
non-radiative recombinations are most effective and dominant (high
temperatures).

Whereas there is general agreement among different experimentalists that $\tau$
strongly depends on temperature,\cite{Calcott1993,Vial1992,Pavesi1993} the
situation is less clear regarding the exponent $\beta$.  While some authors
report values of $\beta\cong 0.8-0.9$, independent of the temperature
\cite{Kondo1993,Kanemi1993}, others found temperature dependent values.
\cite{Ookubo1994,Pavesi1993,Pavesi1995} In this case, the $\beta$  values
follow similar temperature and energy dependences in the range 0.4-0.8.  By
varying the parameters in our model, we are able to describe both
situations.  Temperature dependent $\beta$ values are shown in Fig.
\ref{f-compare3D}.  Temperature independent values can be obtained when
$\tau_{\rm hop} < \tau_{\rm nr} and \tau_{\rm sing}$, since in this case the
activation term in Eq.  \ref{hopp}, which is mainly responsable for the
temperature dependence of $\beta$, does not influence the diffusion of
carriers, being only restricted by the geometrical constraints of the p-Si
network.

By going from an interconnected array of QDs to an ensemble of isolated QDs
our model predicts that the lifetime $\tau $ and the dispersion exponent
$\beta$ increase, with $\beta$ taking values near 1.  This has been indeed
observed in several experiments.  For example, the data presented in Ref.
\onlinecite{Yamada1992}, and reanalised using stretched exponential
functions, show that $\tau $ and $\beta$ increase after dry oxidation of
p-Si.  It is well know that the interconnected array of QDs typical of p-Si
is transformed after dry oxidation into a dispersion of nano-crystals
immersed in a SiO$_x$ matrix; increasing further the oxidation a porous
glass is formed.\cite{Canham1993}

Our simulations show that hopping of excitons is responsable for the
stretched exponential decay observed in p-Si.  This rules out other
explanations which are based on analogies with hydrogenated amorphous Si
(a-Si:H), see e.g.  Ref.  \onlinecite{Tessler1993}.  In fact we
demonstrated that PL decay lineshapes, strongly resembling those measured
in a-Si:H, are found for uncorrelated e-h pairs.  This behavior can
possibly explain the lack of room temperature luminescence in a-Si:H.  In
fact, it is possible that the fast diffusing species in the pair is rapidly
trapped into non-radiative recombination centers when the thermal energy is
sufficiently large to promote the particle to extended
states.\cite{Pavesi1995c}

Annealing treatments induce a quenching of the room temperature
luminescence with a decrease in the values of $\tau $ and
$\beta$.\cite{Ookubo1992} Within our model, this is simulated by
diminishing $\tau^{(0)}_{\rm nr}$ and $\tau_{\rm hop}$ yielding lower
values of $\tau $, $\beta $ and quantum efficiency (see Table
II).  The separate effects of variing $\tau^{(0)}_{\rm nr}$ and
$\tau_{\rm hop}$ are the following:

\noindent (a) Reducing the non-radiative lifetime, yields an increase of
$\beta$ and a decrease of $\tau$.  The reason is that the re-population of
the target QDs, those from which the luminescence decay is recorded, is
reduced by the competitive non-radiative recombination channels.  Less
excitons reach the target QD at long times because a large part recombine
non-radiatively as they diffuse through the system.  The target QD
effectively behaves as being more isolated (increase of $\beta$).  Then,
$\tau $ is reduced due to the competing on-site non-radiative
recombinations, and consequently the quantum efficiency is also reduced.

\noindent (b) Reducing $\tau_{\rm hop}$ yields a reduction of $\tau $ and
$\beta$.  In fact, for higher hopping probabilities, the target QDs are
emptied faster due to the competing loss of excitons caused by hopping
(reducing $\tau$), while at long times more excitons reach the target QD
giving rise to long-lived recombinations (reducing $\beta$).  In this case,
no variations in the quantum efficiency are expected.

\noindent Hence to obtain the trends measured during the annealing
experiments both the lifetimes, $\tau^{(0)}_{\rm nr}$ and $\tau_{\rm hop}$,
should be reduced.

Let us compare our simulations with other models.  The non-exponential
decay of L(t) has been explained as due to a distribution of $\tau_{\rm
rad}$ values arising from a shape distribution of quantum dots with the
same emission energy.\cite{Calcott1993} Within this assumption, however, it
is difficult to explain the temperature dependence of $\beta$ already
discussed.  It has also been proposed that the non-exponential behavior of
L(t) is a consequence of a distribution of non-radiative decay
rates.\cite{Suemoto1994,Sawada1994} This fact is not considered in our
model, and a refined version of it should certainly contain such a
dependence.  However, we want to emphasize here that it is the dispersive
motion of excitons which is mostly responsible for the non-exponential
decay of L(t).  The radiative and non-radiative lifetime distributions
could indeed concur to produce a similar effect on L(t), but in isolated
and well passivated quantum dots it is reasonable to expect essentially a
simple exponential decay of L(t).  Such a behavior has been indeed observed
for some p-Si samples.\cite{Hooft1992}

The strong temperature dependence of the effective anomalous diffusion
exponent $d_w$, and the logarithmic time dependencies of $R^2(t)$ at lower
temperatures, have important consequences on the ac-conductivity, which is
expected to display a strong frequency dependence at low temperatures.
This observation is supported by recent experimental results on the
ac-conductivity of p-Si.\cite{Conduct} In contrast, at higher temperatures,
diffusion becomes normal, $R^2(t)\sim t$, i.e.  $d_w\to 2$, and the
dispersion frequency range shrinks considerably.  It should be emphasized
that a model for the structure of p-Si has been suggested
recently,\cite{Conduct} in which the p-Si network is assumed to be the
infinite percolation above criticality.  From our present results, however,
one can see that such a model is far too simple to describe p-Si samples of
different porosities, and a more elaborated model is required.  In
addition, theoretical results for the ac-conductivity discussed so
far,\cite{Conduct} have been obtained using mean-field approximations, and
the new intrinsic behavior suggested here for diffusion of carriers in p-Si
at low temperatures can not be obtained.

\section{Conclusion}

The theoretical results obtained in this paper have for the moment a
semi-quantitative character.  A quantitative comparison between the present
results and the available experimental data can be made possible when the
parameters entering the model may be estimated independently and more
accurately.  Also, the precise relations between the nano-crystal structure
and dynamical properties are so far unkown.  We have tried to fill this
lackness by making ad hoc, yet standard assumptions which seem to describe
the physics of p-Si rather well.  In addition, the actual values of the
parameters are so strongly sample dependent and influenced by various
treatments (ageing, oxidation, storage, excitation conditions, etc.) that
quantitative predictions valid in general are not possible.  However,
several experimentally measured trends in the photoluminescence lifetimes,
dispersion exponents and quantum efficiencies of the luminescence are
correctly explained in the frame of the present calculations by choosing
the free parameters of the model appropriately.  This gives a strong
support for the validity of the present model for describing the
recombination dynamics in p-Si.

Finally, by assuming different forms for the various quantities reported in
Section II, other systems can be described as well by the present model,
as e.g. nanometer sized Si crystallites,\cite{Kanemi1994}
CdSSe quantum dots,\cite{Dissanayake1995} etc.  All these systems show a
stretched exponential decay of the luminescence.

\newpage
\begin{table}
        \protect\label{table1}
        \caption{ Values used in the Monte-Carlo calculations.
        ${\ell}^{(3/D)}$ is the linear size of the $D$-dimensional
        lattice employed, t$_{\rm max}$ the maximum number of MC
        step for a single run, n$_{\rm runs}$ the number of runs (i. e. the
        number of different particles used) for a single cluster realization,
        n$_{\rm conf}$ the number of different cluster realizations over which
        the reported results are averaged. The meaning of the
        other parameters is given in the text. All the  results presented
        in this paper refer to this choice of parameters unless otherwise
        stated. }
        \begin{tabular}{cccccccccccc}
         $\ell$ & t$_{\rm max}$ & n$_{\rm runs}$ & n$_{\rm conf}$ &
         $\tau_{\rm hop}$ & $\tau^{(0)}_{\rm sing}$ & n$_s$ &
         $\tau^{(0)}_{\rm nr}$ & $\hbar \omega_{\rm ph}/k$  &
         p & Porosity & E$_{\rm cut}$ \\
         31 & 4000 & 2000 & 200 & 150 $\mu$s & 150 $\mu$s & 1.5 &
         900 $\mu$s & 800 K & 25 & 65 \% & 0.04 eV \\
        \end{tabular}
\end{table}

\begin{table}
        \protect\label{table2}
        \caption{300 K simulation results for a 3D clusters of $65\%$
         porosity as a function of the time parameters:
         $\tau^{(0)}_{\rm nr}$ and $\tau_{\rm hop}$. The other parameters
         are given in Table I.}
         \begin{tabular}{ccccc}
          $\tau^{(0)}_{\rm nr}$ & $\tau_{\rm hop}$ & $\tau$ & $\beta$ & QE  \\
          ($\mu$s) & ($\mu$s) & ($\mu$s) & & \%  \\
          \hline
                 900 & 150 & 72 & 0.77 & 12  \\
                 800 & 300 & 84 & 0.87 & 10  \\
                 800 & 100 & 59 & 0.72 & 10  \\
                 500 & 150 & 51 & 0.84 &  7  \\
         \end{tabular}
\end{table}

\newpage
\begin{figure}
\caption{Distribution of optical-band gaps for an ensemble of quantum dots
(dotted line) (Eq.  2), and corresponding density of states (continuous
line) (Eq.  4) versus energy.}
\label{f-Gaus}
\end{figure}

\begin{figure}
\caption{Temperature dependence of the transition times (Eq.  13) for an
energy of 1.86 eV: $\tau_{\rm rad}$ (short-dashed lines), $\tau_{\rm nr}$
(long-dashed lines) and $\tau$ (continuous line).  Here we have used the
same set of parameters as given in Table \protect\ref{table1} except for:
case (1), thick lines: $\tau ^{(0)}_{\rm sing}=150~\mu{\rm s}$, $\tau_{\rm
nr}^{(0)}=900~\mu{\rm s}$; case (2), thin lines: $\tau^{(0)}_{\rm
sing}=100~\mu{\rm s}$, $\tau_{\rm nr}^{(0)}=400~\mu{\rm s}$; and case (3),
thin lines: $\tau_{\rm sing}^{(0)}=200~\mu{\rm s}$, $\tau_{\rm
nr}^{(0)}=7000~\mu{\rm s}$.}
\label{f-taus}
\end{figure}

\begin{figure}
\caption{Quantum efficiency of an ensemble of isolated quantum dots:
$\epsilon(E,T) P(E)$ versus $E$ for various temperatures reported on
the curves in K.  The temperature dependence of the Si band-gap has been
neglected.}
\label{f-qee}
\end{figure}

\begin{figure}
\caption{Total quantum efficiency of the ensemble of quantum dots,
$\epsilon(T)$, versus temperature.  Here we have used the same set of
parameters as given in Table \protect\ref{table1} except for: case (1),
thick line: $\tau_{\rm sing}^{(0)}=150~\mu{\rm s}$, $\tau_{\rm
nr}^{(0)}=900~\mu{\rm s}$; case (2), thin line: $\tau_{\rm
sing}^{(0)}=100~\mu{\rm s}$, $\tau_{\rm nr}^{(0)}=400~\mu{\rm s}$; and case
(3), thin dashed line: $\tau_{\rm sing}^{(0)}=200~\mu{\rm s}$, $\tau_{\rm
nr}^{(0)}=7000~\mu{\rm s}$.  The thick dashed lines represent the
exponential behavior $\epsilon(T)\sim \exp(-T/T_0)$, with $T_0=59$ K for
case (1) and case (2) and $T_0=75$ K for case (3).}
\label{f-qex}
\end{figure}

\begin{figure}
\caption{Absorption coefficient of the ensemble of isolated quantum dots,
         $\alpha(E,T)$, versus energy for the indicated temperatures, T.}
\label{f-abs}
\end{figure}

\begin{figure}
\caption{Snapshot of the energy landscape for diffusion in a
one-dimensional lattice.  The transition energy $E$ is plotted versus
position, for: $E_{\rm cut}=\infty$ (uncorrelated values, top panel),
$0.04$ eV (middle panel), and $0.004$ eV (bottom panel).  The mean energy
value over the 29791 sites is 1.86 eV for the three cases.}
\label{f-1dprofil}
\end{figure}

\begin{figure}
\caption{Monte-Carlo simulation results for a one-dimensional system of
excitons.  Luminescence versus time for four different temperatures: $T=20$
K (squares), 100 K (triangles), 300 K (discs) and 500 K (empty circles).}
\label{f-1dexc}
\end{figure}

\begin{figure}
\caption{Fitting of the luminescence decay (circles) obtained from a
Monte-Carlo simulation for a one-dimensional system of excitons at $T=100$
K.  The following fitting functions and best fit parameters have been used:
Stretched exponential function with $\tau=345$ $\mu$s and $\beta=0.815$
(full line); single exponential decay with $\tau=555$ $\mu$s (dotted line);
double exponential with $\tau_1=80$ $\mu$s and $\tau_2=503$ $\mu$s
(dashed-dotted line).  The top panel reports the relative errors, defined
as the ratio of the difference of the simulation data ($Y$) minus the
computed value ($Y_{FIT}$) times the simulation data, i.  e.
$\left[\left(Y-Y_{FIT}\right)/Y\right]$, for
the three fitting functions.}
\label{f-1dfit}
\end{figure}

\begin{figure}
\caption{Monte-Carlo simulation results for a one-dimensional system of
uncorrelated electron and hole pairs.  Luminescence versus time for four
different temperatures: $T=30$ K (dots), 50 K (circles), 100 K (triangles)
and 300 K (squares).  The lines are power-law fits to the decay with the
following exponents: $\alpha=0.86$ ($T=30$ K), 0.79 ($T=50$ K), 0.73
($T=100$ K) and 0.91 ($T=300$ K).  The curves are vertically shifted for
clarity.}
\label{f-1deh}
\end{figure}

\begin{figure}
\caption{Temperature dependence of the luminescence recombination time
$\tau$ (empty discs) and dispersion exponent $\beta$ (filled discs)
obtained by a stretched exponential fit to the time decay of the
luminescence, for an exciton population in one-dimensional (upper panel),
two-dimensional (middle panel) and three dimensional (lower panel)
clusters.}
\label{f-taubet}
\end{figure}

\begin{figure}
\caption{Mean square displacement $R^2(t)$ of excitons in one dimension
for three different temperatures: $T=20$ K (circles), 100 K (triangles)
 and 500 K (squares). Note the departure from a power-law behavior at low
temperatures. The curves are vertically shifted for clarity.}
\label{f-1dmsd}
\end{figure}

\begin{figure}
\caption{Anomalous diffusion exponent $d_w$ obtained by a power-law fit to
the mean square displacement $R^2(t)$ of excitons in one-dimensional
(circles), two-dimensional (discs) and three-dimensional (triangles) clusters.}
\label{f-dw}
\end{figure}

\begin{figure}
\caption{Energy landscape (transition energies $E$ vs position) in one
dimension.  The arrows are drawn in direct correspondence with the local
energy difference between nearby sites.  The resulting
one-dimensional chain (lower panel) is equivalent to the Sinai model, in
which uncorrelated random fields are present at each site of the lattice.}
\label{f-sinarf}
\end{figure}

\begin{figure}
\caption{Mean square displacement $R^2(t)$ of excitons in one dimension
plotted versus $(\log t)^4$ for the temperatures: $T=20$ K, 30 K and 50 K.
The straight line through the $T=30$ K results represents the asymptotic
$(\log t)^4$ behavior.  The line through the $T=50$ K results shows the
power law $t^{2/d_w}$ behavior, with $d_w=2.5$.}
\label{f-1dlogt}
\end{figure}

\begin{figure}
\caption{Percolation-like clusters grown on a simple cubic lattice
representing porous Silicon of the same 97.4 $\%$ porosity but different
growth probabilities: (a) $p_1=0.9$, $p_2=0.05035$, (b) $p_1=0.3235$,
$p_2=0.4$.}
\label{f-poros3}
\end{figure}

\begin{figure}
\caption{Luminescence decay of a three-dimensional system of excitons
(upper panel) and of uncorrelated electron-hole pairs (lower panel) for
three different temperatures: $T=300$ K (circles), 200 K (discs) and 100 K
(triangles).  The lines are fits to the luminescence decays.  Stretched
exponentials have been used for the exciton case with the following
parameters: $\tau=175$ $\mu$s and $\beta=0.58$ ($T=100$ K), $\tau=135$
$\mu$s and $\beta=0.62$ ($T=200$ K), and $\tau=66$ $\mu$s and $\beta=0.73$
($T=300$ K).  Power laws, $t^{-\alpha}$, have been used for the
uncorrelated electron-hole pairs with the following parameters:
$\alpha=0.96$ ($T=100$ K), $\alpha=1.10$ ($T=200$ K) and $\alpha=1.28$
($T=300$ K).  In this case, the curves are vertically shifted for clarity.}
\label{f-lumi3}
\end{figure}

\begin{figure}
\caption{Temperature dependence of the luminescence recombination times
$\tau$ (empty symbols) and dispersion exponents $\beta$ (filled symbols)
for excitons in three dimensional clusters and for two different sets of
parameters.  Circles correspond to case (1) of Fig.  \protect\ref{f-taus}
and to the standard set of parameters reported in the Table
\protect\ref{table1}.  Triangles correspond to case (2) of Fig.
\protect\ref{f-taus} and hopping time $\tau_{\rm hop}=100$ $\mu$s.  The
lines are the lifetimes calculated using Eq.  \protect\ref{taux} in the
case of isolated quantum dots, i.  e.  without hopping, for case (1) (full
line) and case (2) (dotted line).}
\protect\label{f-compare3D}
\end{figure}

\begin{figure}
\caption{Mean square displacement $R^2(t)$ of excitons in three dimensional
clusters for four different temperatures: $T=20$ K (squares), $T=30$ K
(triangles), $T=50$ K (circles) and $T=100$ K (discs).  The curves are
vertically shifted for clarity.  The lines are power law fits.  Note,
however, the bending of the data at low temperatures for short times.}
\label{f-3dlogt}
\end{figure}

\begin{figure}
\caption{Mean square displacement $R^2(t)$ of excitons in three dimensional
clusters for the temperature $T=30$ K, plotted as a function of $(\log
t)^2$.  Note the linear increase at long times.} \label{f-logt2}
\end{figure}


\begin{thebibliography}{99}
\bibitem{wheni}
A generic quantity X,
representing for instance the particle size $d$, is labeled with an index $i$,
$X_i$, to indicate its actual local value at the  $i$-th site.
\end{thebibliography}
\end{document}